\documentclass[journal, onecolumn]{IEEEtran}

\usepackage[utf8]{inputenc}
\usepackage{amsmath}

\usepackage[T1]{fontenc} 

\usepackage{hyperref} 

\usepackage{mathtools}
\usepackage{graphicx}
\usepackage{tabularx}
\usepackage{subfigure}
\usepackage[table]{xcolor}
\graphicspath{{images/}}
\usepackage{dblfloatfix}
\usepackage{fixltx2e}
\usepackage{multirow}
\usepackage[justification=centering]{caption}

\usepackage{tabularx} 
\usepackage{array} 
\usepackage{booktabs} 

\usepackage{multirow}
\usepackage[table]{xcolor}

\usepackage{amsmath} 
\usepackage{amssymb} 
\usepackage{mathrsfs} 
\usepackage{amsthm} 
\usepackage{mathtools} 

\DeclareMathOperator*{\argmin}{arg\,min}

\begin{document}
\title{Discovering Geo-dependent Stories by Combining Density-based Clustering and Thread-based Aggregation techniques}
\author{H\'{e}ctor Cerezo-Costas, Ana Fernández Vilas, Manuela Mart\'{i}n-Vicente, Rebeca P. Díaz-Redondo
\thanks{H\'{e}ctor Cerezo-Costas (hcerezo@gradiant.org) and Manuela Mart\'{i}n-Vicente (mmartin@gradiant.org) are with Gradiant,  AtlantTIC, Edificio CITEXVI, local 14, Universidade de Vigo, Spain}
\thanks{Rebeca P. Díaz-Redondo (rebeca@det.uvigo.es) and Ana Fernández Vilas (avilas@det.uvigo.es) are with atlanTTic, Universidade de Vigo; Vigo, 36310, Spain.}
}

\maketitle

\begin{abstract}

\noindent Citizens are actively interacting with their surroundings, especially through social media. Not only do shared posts give important information about what is happening (from the users' perspective), but also the metadata linked to these posts offer relevant data, such as the GPS-location in Location-based Social Networks (LBSNs). In this paper we introduce a global analysis of the geo-tagged posts in social media which supports (i) the detection of unexpected behavior in the city and (ii) the analysis of the posts to infer what is happening. The former is obtained by applying density-based clustering techniques, whereas the latter is consequence of applying natural language processing. We have applied our methodology to a dataset obtained from Instagram activity in New York City for seven months obtaining promising results. The developed algorithms require very low resources, being able to  analyze millions of data-points in commodity hardware in less than one hour without applying complex parallelization techniques. Furthermore, the solution can be easily adapted to other geo-tagged data sources without extra effort.

\end{abstract}

\begin{IEEEkeywords}
data mining, crowd detection, density-Based clustering, content aggregation, event detection
\end{IEEEkeywords}

\section{Introduction} \label{Introduction}

Nowadays, users are the main source of alternative sensor information in a city, although this huge source of information is often overlooked. Being ubiquitously connected to the Internet with their mobile phones, they intensively use services which promote user generated content such us Online Social Networks (OSNs), one of the most massively alternatives employed. Content in OSNs is a combination of text/images (e.g. a user post, a reply to other users posts, etc.) and meta-data information (number of likes, stars of user posts, number of posts made by the user, GPS-location, etc.). When using a GPS-enabled device, users also add a very valuable information: from where the post is shared. Thus, by analyzing the geo-located posts it is possible to know what is happening and where it is happening \cite{adedoyin2016rule, hua2016automatic}. This is especially relevant in OSNs adapted for fast consumption (e.g. microblogging or image messaging) in which the time lapse between an event and its appearance in the platform is very low. 

Our previous work \cite{dominguez2017sensing} introduces an approach to take advantage of the information given by the geo-located posts shared in social media. Abnormal location patterns were detected in the urban area under study, such as unusual city states or dynamics. The input data of the model was restricted to the posts' geolocation. This information was employed in order to find out when the shared posts in a specific area at that time of the day and that day of the week can be considered usual or an outlier (too much or too many). For this to be possible, a density-based clustering technique was applied. After a training stage which obtained the usual pulse of the city, the technique allowed the detection of abnormal behaviors on-the-fly. The work introduced in this paper supplements our previous findings. Here we set up a two-folded approach. Once the abnormal location pattern is detected, it identifies what is going on, where and when. Taking  the set of posts which lead to a geo-anomaly as an activity  seed, our new proposal enlarges the focus to all those posts which are considered linked to the seed. Opposite to pure NLP (Natural Language Processing) for geo-dependent topic modelling \cite{CAPDEVILA201758, Xia2015}, we apply Content Aggregation Models as the one in \cite{Hodson:2015:2050-0076:249} to identify meaningful threads of content that reflect what is happening in the area under study in a timely fashion. We do this in order to react to potential threats as soon as possible.

The paper is organized as follows. Section \ref{sec:Relwork} summarizes other research proposal that are relevant for our work. Section \ref{sec:methodology} overviews the proposed methodology of the events detection system. In Section \ref{sec:dataset} we describe the dataset and the reasons behind our selection of Instagram as data source. Section \ref{sec:Evaluation} details the main aspects that have focused the evaluation of our proposal, whereas in Section \ref{sec:Results} we enumerate the obtained results after the different experiments performed. The results are discussed in Section \ref{sec:Discussion} and, finally, in Section \ref{sec:Conclusions} we outline the conclusions and future work.

\section{Related Work} 
\label{sec:Relwork}

Analyses of data gathered from social media (text and location linked to geo-tagged posts) have been recently applied for different and interesting purposes related to mobility patterns. In \cite{Hawelka:2014} for instance, a worldwide analysis of travelers is performed by using geo-located tweets. The approach was validated by comparing the results to global tourism information, showing a strong correlation. This travelers flow enables the detection of different communities in different countries, reflecting a regional division of the world. Another interesting approach is introduced in \cite{Frank:2013}, where sentiment analysis techniques are applied to about 180,000 geo-located tweets to infer the relationship between happiness and movements within a city.

In this paper, we focus our attention to another interesting field: the detection of crowds and events in urban areas. With this aim, information gathered from shared posts supports the application of different analysis techniques applied to both the text and the location of the geo-tagged posts.

\subsection{Crowds and events detection}
\label{sec:related-crowds}

With an approach which constructs clusters of tweets according to their number in a given area (density),  the detection of local events is the main aim in \cite{walther2013geo}. Afterwards, these clusters are scored according to different criteria: textual content, number of users, number of tweets, etc. Quite similarly, the authors in \cite{dokuz2017discovering} developed a customized density algorithm to obtain socially  interesting locations in a city using geo-located tweets. In the first stage, they obtain the prevalence of locations for each user. After that, interesting locations, from the poin of view of group behavior,  are discovered combining the per-user results. In \cite{ranneries2016wisdom} posts from both Twitter and Instagram are clustered according to their hashtags. After that, the density-based clustering algorithm DBSCAN is applied to these clusters in order to associate a single place to each cluster. A different clustering approach is presented in \cite{lee2010measuring}, where k-means is used to group the geolocated tweets and define Regions of Interest (RoI). Over these regions, the number of tweets is analysed in order to detect outliers. The objective is to develop a geo-social event detection to monitor crowd behaviors and local events. The approach introduced in \cite{lee2012mining} tries to infer spatio-temporal information about the events mentioned in the shared tweets. Authors applied text mining techniques (a density-based online clustering method), with the aim of detecting events in urban areas. In this approach, the location of an event is extracted direclty from the text content when the geo-tagged information linked to the tweets is not available.

In \cite{ferrari2011extracting}, most visited locations are detected applying the EM-Algorithm to the location of tweets in intervals of two hours. These popular places are associated to a ZIP code. Those ZIP codes are processed using Latent Dirichlet Allocation (LDA) to find patterns in the movements of the crowds and track events with a strong relation with the city. LDA is also applied in \cite{chae2012spatiotemporal}, in this case to the text content of the tweets, in order to find popular topics. Then an abnormality estimation is calculated using Seasonal Trend Decomposition based on Loess smoothing (STL), in an iterative process which requires expert human supervision. Other LDA-based approach is detailed in \cite{Yuan:2012} to relate topics and regions. Once the topics are obtained, a clustering technique is used to aggregate regions with similar topic distributions. Topic distribution is also the base of the approach in \cite{Watanabe:2011}, where local events are detected from  analyzing microblogging data. Geohash application \footnote{http://geohash.org} is used for clustering location data and authors. They apply keyword frequency as the discriminatory factor for aggregating content. Keywords are associated with regions when both appear jointly more than three times in the dataset to identify local events in a region. Although the simplicity of this approach is suitable for online analysis, the event extraction schema is too naive to provide the filtering capabilities needed for anomaly incident detection.

\subsubsection{Clustering and outlier techniques}
\label{sect:backgroundClustering}

There are multiple clustering methods, which are generally classified in four groups: partitioning approaches (where the number of clusters is pre-assigned), grid-based (where the object space is divided into a pre-assigned number of cells), hierarchical (where the data is organized in multiple levels) or density-based (where density notion is considered). For our purpose, density-based algorithms are the most suitable since they are able to (i) discover clusters of arbitrary shapes, (ii) handle sparse regions (which are considered as noisy regions) and (iii) work without knowing the number of clusters in advance. Among the different proposals in the literature \cite{dbscancomparative}, we selected DBSCAN~\cite{dbscan} (Density-Based Spatial Clustering of Applications with Noise). Two parameters are necessary in DBSCAN to define the density measure to  obtain the clusters: the radius of a circle around the data point ($\epsilon$) and the minimum number of points that should be in this circle in order to be considered a cluster ($minPoints$). The algorithm is very sensitive to both parameters, so it is essential to select their values properly. Our estimation algorithm, detailed in~\cite{dominguez2017sensing}, is adaptive (since it is based on the nature of the dataset) and has less time complexity than other approaches in the literature.  

After being able to detect groups of geographically close citizens with activity in social media (crowds) by using DBSCAN, the second step is defining the conditions under which these crowds are considered outliers. According to Hawkins "{\em an outlier is an observation that deviates so much from other observations as to arouse suspicion that it was generated by a different mechanism}"\cite{outlier}. As described in ~\cite{dominguez2017sensing}, we treat a cluster as an outlier  whenever the number of points differs from the number of points of other clusters found in a similar location, day and hour. We also differentiate between mild and extreme outliers \cite{boxplot}: the former lies outside the interval  $(Q_1 - 1.5IQR, Q_3 + 1.5IQR)$, whereas the latter lies outside  the interval $(Q_1 - 3IQR, Q_3 + 3IQR)$, being $IQR=Q_3 - Q_1$ the Interquartile Range.

\subsection{Content Aggregation Models}
\label{sec:related-content}

Content aggregation usually involves one-to-one similarity comparisons of records (with $O(n^2)$ complexity). In applications that operate in the order of millions, a greedy comparison strategy is unmanageable. Hence, clustering techniques, such as the aforementioned LDA \cite{Blei:2003}, reduce the number of required comparisons. However, LDA presents two characteristics that determine its scope. Not only the number of clusters has to be previously specified but also topics should be extracted over historic data, a time consuming task. Other alternatives rely on Bayesian non-parametric models such as DPMM \cite{antoniak1974mixtures} to obtain topics without supervision \cite{lo2017unsupervised}. One important advantage of DPMM over LDA is that clusters must not be predefined. This is of paramount importance in social media due to the dynamism of the content in these services. 

Locality Sensitive Hashing (\textbf{LSH}) \cite{Indyk:1998, Gionis:1999, Datar:2004} is an online clustering technique with application in $\mathbb{R}^n$. Although it is a general purpose clustering technique, it has interesting applications in text analysis, which can take advantage of its speed and dimensionality reduction. LSH uses random projected vectors (hypervectors) to split the space in buckets (Figure~\ref{fig:lsh1}). These buckets are organized in such a way that vectors closer in the original $\mathbb{R}^n$ have more chances of being in the same bucket that those which are further apart. Therefore, those buckets work as a filtering technique to drastically reduce the number of comparisons performed among records (Figure~\ref{fig:lsh2}). The complexity of the model is thus characterized by the number of hypervectors, the original dimensionality of the space, and the records that lie in each bucket. There is a tradeoff between precision and complexity. The number of buckets increases with the number of hypervectors which as a consequence will lower the number of records stored in a bucket. Nevertheless, the system will be less precise since the chance of two points being separated by a hypervector increases as well.

LSH has been applied in \cite{Petrovic:2010} to find the points closest in space and to build threads of stories of an OSN (Twitter). They use the TF-IDF \cite{Ramos:2003} algorithm to obtain the set of characteristics of the input vector, but they strip the system of hashtags. As they apply word-level characteristics, the system must fix beforehand the words that might be used as vector components in the input. For this reason, they remove hashtags and user mentions despite being extremely relevant information for content aggregation in social media. 

\section{Methodology: an Overview}
\label{sec:methodology}

\begin{figure*}
    \centering
    \includegraphics[width=0.95\textwidth]{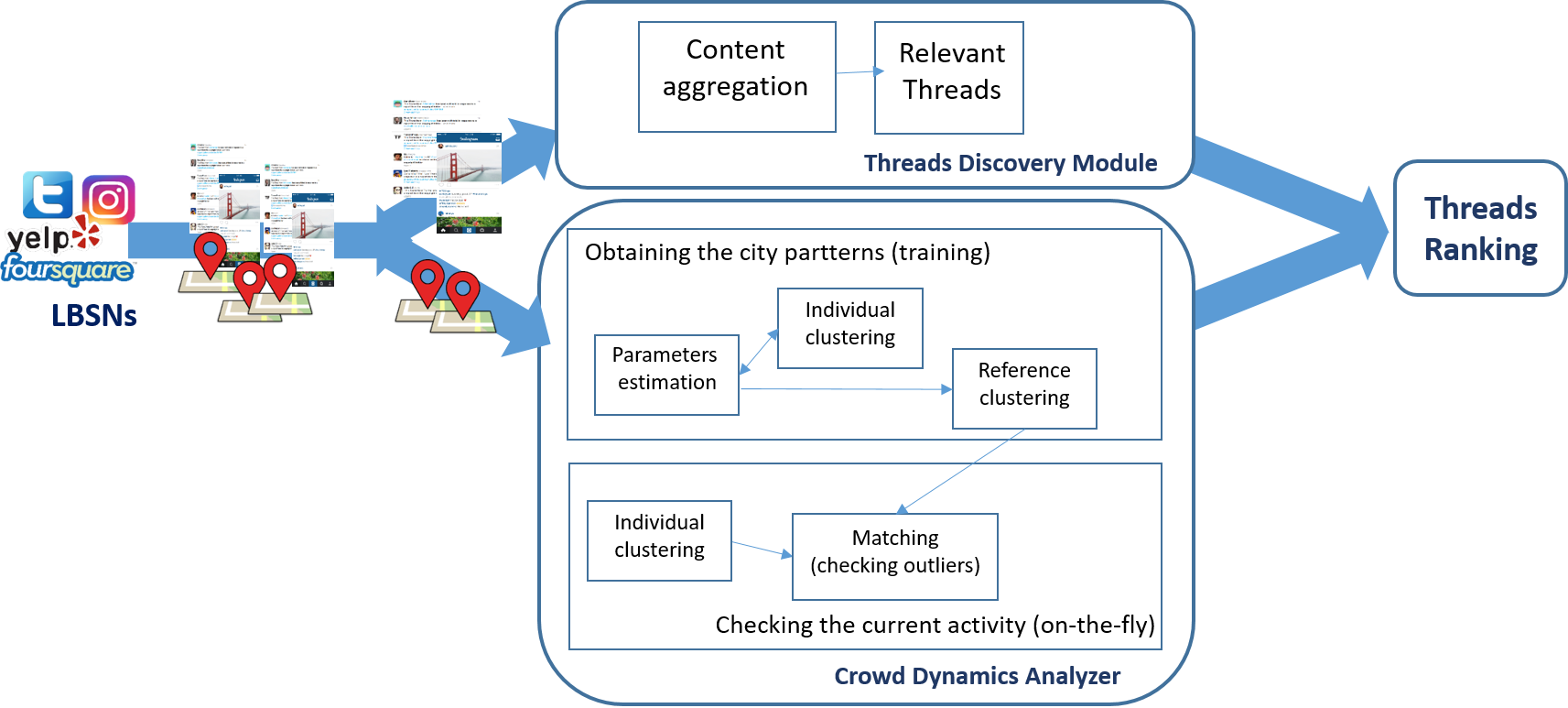}
    \caption{Events Detection System}
    \label{fig:archearlydet}
\end{figure*}

Our approach combines the analysis of two kinds of data obtained from the same source (LBSNs): text and GPS location of the shared posts (Figure~\ref{fig:archearlydet}). Whereas location information is forwarded to the Crowd Dynamics Analyzer module, text is the input of the Threads Discovery module. Both modules work independently to detect unexpected citizens' activity in social media, assuming that this behavior reflects that something unusual is happening in the area. The Crowd Dynamics Analyzer (Section~\ref{sec:crowds}) focuses on the GPS-data of the posted messages to check if the locations are the usual ones for the combination of day of the week, time and area. The Threads Discovery Module (Section~\ref{sec:threads}) deals with the content (text) of the posted messages to infer sequences of posts belonging to the same theme. Finally, both modules feed the Threads Ranking (Section~\ref{sec:ranking}), whose main aim is to check if the detected threads effectively correspond to an unusual crowd behavior. 

\subsection{Crowds Dynamics Analyzer}
\label{sec:crowds}

As aforementioned, this module is in charge of detecting unexpected behaviors in the urban area under study by analysing the citizens' activity in social media. Our approach firstly infers the activity pattern of the city, i.e. the usual behavior (number of shared posts). This behavior varies according to three different parameters: time of the day, day of the week and geographic area within the city. Thus, for instance, the usual activity on Saturday night at the Meatpacking District is totally different to the usual activity at the Financial District. Exactly the same happens if we select another day and time (Monday morning, e.g.). This is, in fact, the main purpose of the first stage of our methodology~\cite{dominguez2017sensing}, the Training Phase, and one of the aspects that differentiate our approach from others in the literature (Section~\ref{sec:related-crowds}), where no patterns are provided.

In order to obtain these patterns, we combine data from several days that can be considered similar (same day of the week in our approach) and then we analyze the social media activity in slots of 30 minutes. Thus, we have 7 reference days (from Monday to Sunday) and 48 time slots per reference day. The procedure can be described as follows: (i) we firstly apply the DBSCAN algorithm to the data (geo-located info linked to each post) obtained for each temporal slot and for each day. Thus we are able to  identify clusters or crowds (i.e. closer locations) and discard noise; (ii) then, we apply the DBSCAN algorithm to the data of all days together with the aim of identifying reference clusters or crowds in the city and (iii) we finally specify two thresholds to detect moderate outliers and extreme outliers. Therefore, after this first Training Phase, we finally provide different patterns (one per combination of day of the week and time slot) outlining where the crowds are located (detected clusters) and specific measures to detect outliers. More details can be found in~\cite{dominguez2017sensing}.

Once the patterns of the city are available, it is time to run the Detection Phase. This is performed on-the-fly: i) gathering the data from LBSN; (ii) clustering the data using the same parameters obtained in the Training stage; and (iii) analyzing if there are outliers or not.

In this phase, it is essential to specify a measure of distance to match individual clusters obtained in the Detection Phase, $C_x$, with the reference clusters defined in the pattern of the city, $P_y$. We have defined this distance, $dist_{xy}$, as follows:

\[dist_{xy} = \frac{1}{n_{C_x}} \sum_{i=0}^{n_{C_x}} dist_{x_{iy}}\]

\noindent were $n_{Cx}$ is the number of points in the cluster $C_x$ and $dist_{x_{iy}}$ is the  distance between the point $c_{x_i}$ (point $i$ in cluster $C_x$) and the nearest point $p_{y_j}$ in the Reference Cluster $P_y$:

\[dist_{x_{iy}} = min(dist(c_{xi},p_{xj})), \forall p_{y_j} \in P_y\]

\noindent We apply the Haversine distance to take into account the fact that the points are
on the surface of the Earth. Finally, we consider the cluster $C_x$ to fit Reference Cluster $P_y$ if it holds that:

\[P_y = \argmin(dist_{xy}) / dist_{xy} \leq \epsilon\]
 the 1\% of the published tweets and only 3.17\% of those were geo-located tweets. On the other hand, although Twitter Search API is also free-access and designed to collect old tweets, they limit the number of calls both per user (180 calls/15 minutes) and per application (450 calls/15 minutes). Foursquare had also a main problem: posts are always linked to the venues in which users share their opinions and comments. Therefore, posts location is biased by the venues location. Finally, Instagram offered several advantages: (i) it did not impose any restriction to the location linked to posts, it being directly the users' GPS-location; (ii) it was possible to directly recover all the posts shared within a specific geographic area; (iii) due to its growth in the recent years \cite{duggan2015social}, Instagram already had more monthly active users than Twitter (in January 2016), and a higher number of the posts contain information about location; and (iv) finally, Instagram API went also further than Twitter: (a) allowed gathering posts published at any moment in the past and (b) imposed less calls restrictions (500 calls/hour for Sandbox mode, 5000 calls/hour for Live mode).

According to the aforementioned reasons, Instagram was selected to populate our NYC data set by gathering data from a circular area (5 Km.) centered in Times Square (40.756667 N, 73.986389 W) from 23rd August 2015 to 28th February 2016 (190 days): $4,335,880$ posts. For this experiment we focused our attention on Saturdays, $783,907$ posts, which were grouped in chunks of 30 min. This long period covered: (i) special days when the city is traditionally more crowded like Christmas time; (ii) unusual days, like the weekend when Storm Jonas hit the United States, when the city was less crowded; (iii) days which are considered normal, when no special events or phenomena are expected to happen and (iv) days during which some events happen with high impact in small areas for a short period of time, like the New York Comic-Con.

\begin{figure}[!htbp]
 \centering
 \subfigure[Times Square \label{fig:times}]{\includegraphics[width=0.48\textwidth]{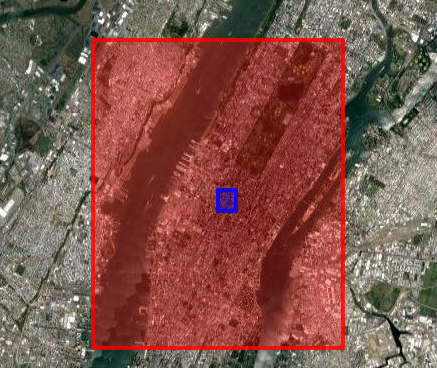}}
 \subfigure[NY Comic-Con \label{fig:comic}]{\includegraphics[width=0.48\textwidth]{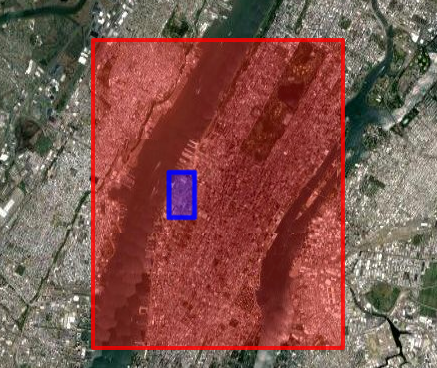}}
  \caption{Filtered zone within the area of analysis.}
 \label{fig:}
 \end{figure}

\section{Evaluation} 
\label{sec:Evaluation}

In order to validate the Events Detection System, we performed different experiments. Firstly,  we proceed only in delimited areas where well known events were celebrated to manually check how the behavior of the system was, i.e. the content threads extracted. The intention was mainly to check the consistency of the methodology. Then, we checked the whole processing pipeline, i.e the Crowds Dynamics Analyzer, the Thread Discovery module and the Event Ranking module. The objective here was to confirm the filtering capabilities of the whole system to be used for automatic incident detection. In these experiments, content threads were obtained using the whole dataset. Finally, after running the system without previous restrictions, not only  all relevant events were detected, but also new ones that were overlooked during the manual review. Among all the experiments that were performed, we highlight four because of their specific characteristics.

\paragraph{New Year's Eve}
New Year's Eve is a highly interesting data since an unusual amount of people concentrates all around the city for a short period of time (a few hours). In our case, we focus our attention on Times Square and its surroundings (Figure~\ref{fig:times}) to analyze the data gathered from this area. Finally, we compared these results with the analysis performed on the data obtained out of this area. As a brief summary, a total number of $47,617$ posts were analyzed. Although the area next to Times Square is small, a significant percentage of the posts came from there, around $5\%$.

\paragraph{Comic-Con}
NY Comic-Con was interesting because an unusual amount of people concentrates in a relatively small area. After checking the venue, we restricted the area under analysis as Figure~\ref{fig:comic} shows and the time to the weekend where the event took place. As a brief summary, the sub-area of influence of the Comic-Con contained all the posts about this topic, having analyzed all the posts shared in 2015/10/10 ($23,545$).

\paragraph{Storm Jonas} 
Lastly, Storm Jonas hit NYC and literally paralyzed the urban area from 21st to 24th of January 2016. Our interest here lies in the analysis of a totally opposite behavior: abnormally low activity level all around the city and potentially located in different areas than it is usual. We checked if the Crowds Activity Analyzer answered as expected and if the Threads Discovery Module was able to infer relevant threads of content related to this event, despite of the large influence area. In this case, we analyzed a dataset of $25.355$ posts.

\paragraph{Unexpected Events} in order to check the ability of the system in detecting geolocated events when no prior knowledge exists, we use as input all the Saturday posts from 2015/08/29 00:00 to  2016/01/23 23:30 in the area of analysis (around 800 mil posts).

\paragraph{Technical restrictions}
Finally, it is important to remark that all the operations were carried in commodity hardware (Intel(R) Core(TM) i5-4430S CPU @ 2.70GHz) with 16GBytes of RAM. The Content Thread generation module was able to work at a pace of 250 posts/s. Memory is also constrained because of the characteristics of the algorithm employed (it depends on the number of spaces, hypervectors, internal temporal memory and bucket size).

\section{Results} 
\label{sec:Results}

As aforementioned, the experiments were oriented to check two main aspects: (i) the Threads Discovery Module and (ii) the whole Events Detection System. The performance of the Crowds Dynamics Analyzer module was already assessed in detail in~\cite{dominguez2017sensing}, having obtained very good results.

\subsection{Threads Discovery Module}
\label{sec:resultsThreads}

In order to assess this module we focused our attention on two events with local influence: the New Year's Eve and the NY Comic-Con. After analyzing the datasets specified in Section~\ref{sec:Evaluation}, we compared the obtained threads with the topic of each event.

Regarding New Year's Eve the most relevant threads identified by our system are summarized in Table \ref{threads:nyeve:in} and Table \ref{threads:nyeve:out}, comparing respectively the results within and out of the restricted area. Threads detected within the Times Square Area mention the location where the celebration takes place. In contrast, outside the Times Square area, the relevant threads mention other important landmarks of the city (such as Central Park) or mentioned only the event (New Year) without making reference to any specific location. Threads are calculated for different similarity thresholds. Despite outputs differ slightly in terms of content aggregation, the relevant information is displayed with all the considered configurations.

\begin{table*}[htb]
\begin{center}
\begin{tabular}{|c|l|c|}
\hline \bf Threshold & \bf Story Threads & \bf Size \\  \hline

\multirow{3}{*}{0.6} & \parbox[t]{8cm}{\textit{Happy New Year!!!! \#2016 \#happynewyear \#timesquareballdrop \#timesquarenewyear }} & 819 \\ \cline{2-3}
 & \parbox[t]{8cm}{\textit{The day after The Ball dropped \#timesquare \#newyear \#newyork }} &  155 \\ \cline{2-3}
 & \parbox[t]{8cm}{\textit{City  is dead right now (sleeping city) \#newyork \#nyc \#manhattan }} & 44\\
 \hline
 
\multirow{3}{*}{0.65} & \parbox[t]{8cm}{\textit{Happy New Year 2016!!! \#newyearseve \#timessquare \#2016 \#nyc}} & 498 \\ \cline{2-3}
 & \parbox[t]{8cm}{\textit{Amazing night! \#happynewyear \#timessqueare \#balldrop \#countdown}} &  156\\ \cline{2-3}
 & \parbox[t]{8cm}{\textit{Wishing everyone a safe and wonderful New Years Eve 2016 tonight ! ...}} & 41\\
 \hline 
  
 \multirow{3}{*}{0.7} & \parbox[t]{8cm}{\textit{\#timessquare \#timesquare \#ny \#newyork \#nyc \#happynewyear2016}} & 368 \\ \cline{2-3}
 & \parbox[t]{8cm}{\textit{One hour to go! \#NYE \#timessquare \#balldrop}} &  97\\ \cline{2-3}
 & \parbox[t]{8cm}{\textit{We did it \#timessquare \#newyearseve}} & 26\\
 \hline 
 
 \multirow{3}{*}{0.75} & \parbox[t]{8cm}{\textit{\#timessquare \#timesquare \#ny \#newyork \#nyc \#happynewyear2016}} & 143 \\ \cline{2-3}
 & \parbox[t]{8cm}{\textit{\#ny \#newyork \#manhattan \#nyc \#broadway \#usa \#timessquare}} &  167\\ \cline{2-3}
 & \parbox[t]{8cm}{\textit{\#rooftop \#timessquare \#balldrop \#newyears}} & 76\\
 \hline 
 
\hline
\end{tabular}
\end{center}
\caption{\label{threads:nyeve:in} Top-three threads under the area of influence of the Times Square for different threshold values}
\end{table*}

\begin{table*}[htb]
\begin{center}
\begin{tabular}{|c|l|c|}
\hline \bf Threshold & \bf Story Threads & \bf Size \\  \hline

\multirow{3}{*}{0.6} & \parbox[t]{8cm}{\textit{Banning sweets is clearly not on my New Year's resolutions list \#food \#foodporn \#dessert \#dessertporn}} & 2656 \\ \cline{2-3}
 & \parbox[t]{8cm}{\textit{Happy new year \#nyc \#newyork}} &  2545\\ \cline{2-3}
 & \parbox[t]{8cm}{\textit{\#happynewyear}} & 1030\\ \cline{2-3}
 \hline 
 
\multirow{3}{*}{0.65} & \parbox[t]{8cm}{\textit{Happy new year from NYC!}} & 1804 \\
 & \parbox[t]{8cm}{\textit{Wishing you the best of health, wealth, and meaningful times with those you love and care for. Bye-bye 2015!}} &  1528\\ \cline{2-3}
 & \parbox[t]{8cm}{\textit{\#myhappynewyear}} & 761\\ \cline{2-3}
 \hline 
  
 \multirow{3}{*}{0.7} & \parbox[t]{8cm}{\textit{Happy new year from NYC!}} & 1394 \\ \cline{2-3}
 & \parbox[t]{8cm}{\textit{ We would love to take this opportunity to a wish a pleasant prosperous New Year...}} &  869\\ \cline{2-3}
 & \parbox[t]{8cm}{\textit{\#happynewyear}} & 532\\
 \hline 
 
 \multirow{3}{*}{0.75} & \parbox[t]{8cm}{\textit{Happy new year from nyc!}} & 1141\\ \cline{2-3}
 & \parbox[t]{8cm}{\textit{May all your wishes come true. I once had one shoe now I have plenty these are ...}} &  441\\ \cline{2-3}
 & \parbox[t]{8cm}{\textit{\#happynewyear}} & 419\\
 \hline 
 
\hline
\end{tabular}
\end{center}
\caption{\label{threads:nyeve:out} Top-three threads outside the area of influence of the Times Square for different threshold values}
\end{table*}

Similar results were obtained for the NY Comic-Con experiment. The Threads Discovery Module is able to extract meaningful content threads with information about the conference within the restricted area (Table~\ref{threads:comic:in}), whereas the detected threads out of the restricted area do not make reference to the location at all (Table~\ref{threads:comic:out}). 

Finally, and with the aim of analyzing the influence of the similarity threshold (Section~\ref{sec:threads}), we also performed the very same experiments but changing its value ($0.6$, $0.65$, $0.7$ and $0.75$). The results, as Table~\ref{threads:comic:in} and Table~\ref{threads:comic:out} show, were consistent, regardless of the selected threshold. Despite of these good results, we decided to check the effect of the similarity threshold over the whole dataset. The results are depicted in Table~\ref{thread_stats} by showing the number of identified threads. Clearly, the less restrictive the threshold is, the higher aggregation is exhibited by the model. Therefore, and unless specified, we decided to use a similarity threshold of $0.65$.

\begin{table*}[htb]
\begin{center}
\begin{tabular}{|c|l|c|}
\hline \bf Threshold & \bf Story Threads & \bf Size \\  \hline

\multirow{3}{*}{0.6} & \parbox[t]{8cm}{\textit{Found Xur before the weekend ended along with Eris. \#destiny \#xur \#nycc \#nycc2015 \#newyorkcomiccon}} & 324 \\ \cline{2-3}
 & \parbox[t]{8cm}{\textit{\#Gaming \#NYCC2015 \#COMICCON}} &  320\\ \cline{2-3}
 & \parbox[t]{8cm}{\textit{Chewbacca \#starwars \#NYCC \#NYCC2015}} & 194\\
 \hline 
 
\multirow{3}{*}{0.65} & \parbox[t]{8cm}{\textit{\#Gaming \#NYCC2015 \#COMICCON}} & 434 \\ \cline{2-3}
 & \parbox[t]{8cm}{\textit{Chewbacca \#starwars \#NYCC \#NYCC2015}} &  162\\ \cline{2-3}
 & \parbox[t]{8cm}{\textit{Comic 411 at Day 2 of \#nycc@newyorkcomiccon \#fan220 @fan220dotcom \#comic411Photo taken by @sonnysofrito}} & 41\\
 \hline 
  
 \multirow{3}{*}{0.7} & \parbox[t]{8cm}{\textit{There goes another Clark lol  \#ComicCon ...}} & 179 \\ \cline{2-3}
 & \parbox[t]{8cm}{\textit{\#Gaming \#NYCC2015 \#COMICCON}} &  154\\ \cline{2-3}
 & \parbox[t]{8cm}{\textit{Chewbacca \#starwars \#NYCC \#NYCC2015}} & 145\\
 \hline 
 
 \multirow{3}{*}{0.75} & \parbox[t]{8cm}{\textit{\#Gaming \#NYCC2015 \#COMICCON}} & 440 \\ \cline{2-3}
 & \parbox[t]{8cm}{\textit{Chewbacca \#starwars \#NYCC \#NYCC2015}} &  167\\ \cline{2-3}
 & \parbox[t]{8cm}{\textit{Comic 411 at Day 2 of \#nycc@newyorkcomiccon \#fan220 @fan220dotcom \#comic411Photo taken by @sonnysofrito}} & 41\\
 \hline 
 
\hline
\end{tabular}
\end{center}
\caption{\label{threads:comic:in} Top-three threads under the area of influence of the NY Comic-Con for different threshold values}
\end{table*}

\begin{table*}[htb]
\begin{center}
\begin{tabular}{|c|l|c|}
\hline \bf Threshold & \bf Story Threads & \bf Size \\  \hline

\multirow{3}{*}{0.6} & \parbox[t]{8cm}{\textit{ swear NYC was fucking lit shoutout to all the Family ...}} & 232 \\ \cline{2-3}
 & \parbox[t]{8cm}{\textit{Cosas que se ven en el central park !! \#newyorker \#centralpark}} &  168\\ \cline{2-3}
 & \parbox[t]{8cm}{\textit{\#newyork \#timessquare \#usa \#traveler \#travel \#voyage}} & 162\\
 \hline 
 
\multirow{3}{*}{0.65} & \parbox[t]{8cm}{\textit{Cosas que se ven en el central park !! \#newyorker \#centralpark}} & 126 \\ \cline{2-3}
 & \parbox[t]{8cm}{\textit{\#timessquare \#newyork \#ny \#travel}} &  114\\
 & \parbox[t]{8cm}{\textit{New York City}} & 111\\ \cline{2-3}
 \hline 
  
 \multirow{3}{*}{0.7} & \parbox[t]{8cm}{\textit{Cosas que se ven en el central park !! \#newyorker \#centralpark}} & 97 \\ \cline{2-3}
 & \parbox[t]{8cm}{\textit{NYC}} &  84\\ \cline{2-3}
 & \parbox[t]{8cm}{\textit{\#timessquare \#nyc}} & 81\\
 \hline 
 
 \multirow{3}{*}{0.75} & \parbox[t]{8cm}{\textit{Cosas que se ven en el central park !! \#newyorker \#centralpark}} & 125 \\ \cline{2-3}
 & \parbox[t]{8cm}{\textit{New York City}} &  121\\ \cline{2-3}
 & \parbox[t]{8cm}{\textit{NYC}} & 111\\
 \hline 
 
\hline
\end{tabular}
\end{center}
\caption{\label{threads:comic:out} Top-three threads out of the area of influence of the NY Comic-Con for different threshold values}
\end{table*}

\begin{table}[h]
\begin{center}
\begin{tabular}{|l|c|c|}
\hline \bf Threshold & \bf Num. Threads & \bf Max. Thread \\  \hline
$0.6$ & $558,208$ & $16,882$\\
$0.65$ & $608,492$ & $7,213$\\
$0.7$ & $647,896$ & $5,365$\\
$0.75$ & $678,551$ & $3,786$\\
\hline
\end{tabular}
\end{center}
\caption{\label{thread_stats} Thread statistics for different similarity thresholds.}
\end{table}

\subsection{Events Detection System}
\label{sec:resultsOverall}

\begin{figure}[!htbp]
 \centering
 \subfigure[NY Comic-Con \label{fig:comiccon}]{\includegraphics[width=0.48\textwidth]{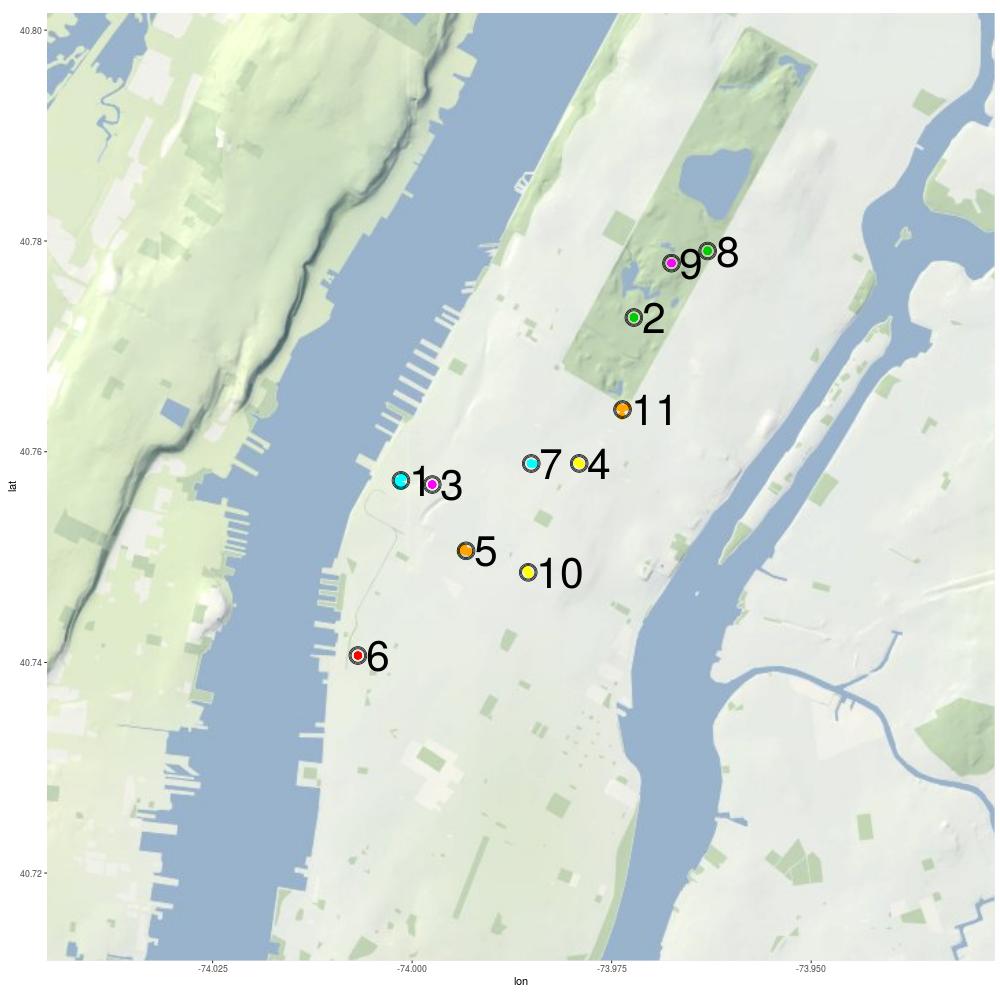}}
 \subfigure[Reference clusters \label{fig:random}]{\includegraphics[width=0.48\textwidth]{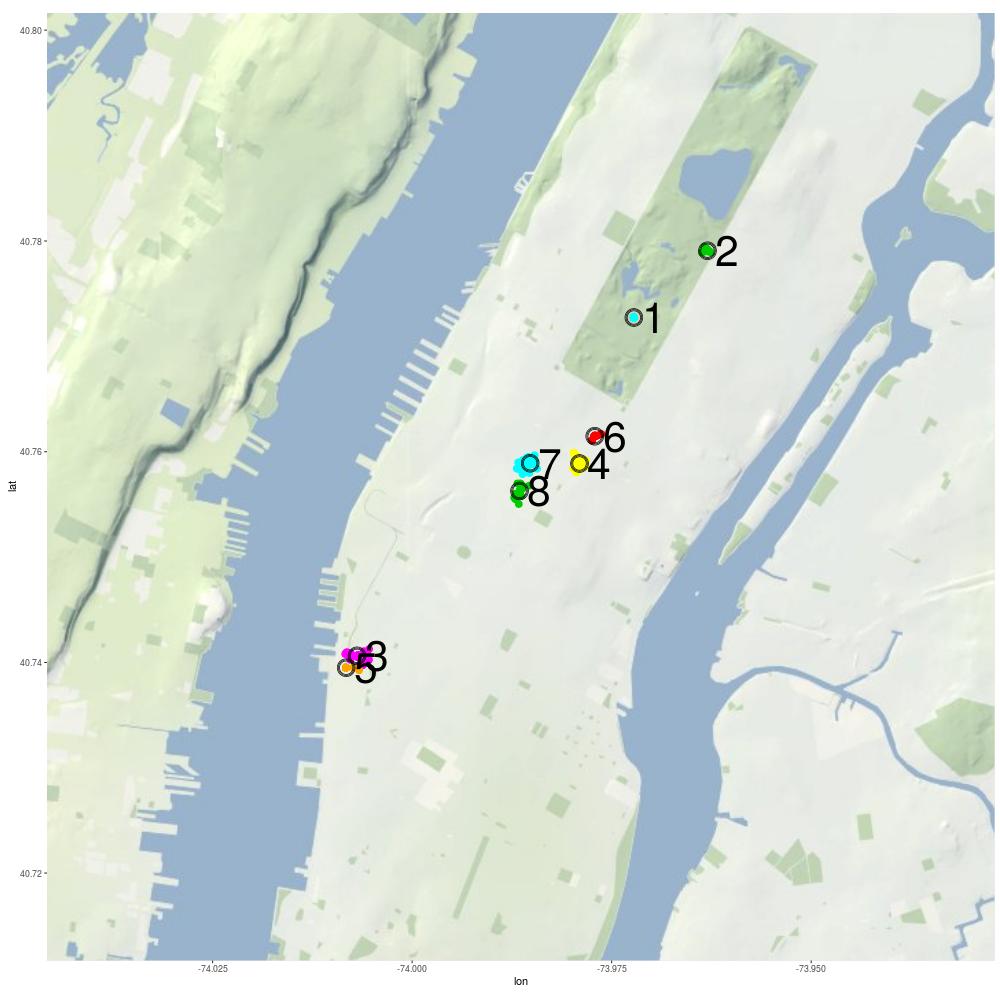}}
  \caption{Geo-located clusters (17:30-18:00)}
 \label{fig:individual}
 \end{figure}

To evaluate the effectiveness of both unsupervised approaches (content and geolocation analysis) we picked up a specific time window (from 17:30 to 18:00) in the NY Comic-Con day to manually assess the top detected stories in this time slot for each geo-located cluster or crowd. On the one hand, the Crowds Dynamics Analyzer detected eleven crowds all around the city that clearly do not fit with the usual clustering or reference clustering of a Saturday at that time (Figure~\ref{fig:individual}).

On another hand, the Threads Discovery Module detected a cluster (number $1$) with 74 posts, whose top threads are detailed in Table \ref{threads:geocluster}. Additionally to this analysis, we have also checked what happens with all the posts that are already shared in Instagram, but that are not geo-located in any cluster ($308$), again the top threads are detailed in the same table. As it can be inferred from the results, the collection of posts which are not included in any cluster is a mishmash of stories within different topics (tourism, a new fossil discovery, etc.), whereas in cluster $1$ stories are strictly delimited to NY Comic-Con related information.

\begin{table}[h]
\begin{center}
\begin{tabular}{|c|l|}
\hline \bf Points & \bf Story Threads \\  \hline

\multirow{4}{*}{non-clustered} & \parbox[t]{6cm}{\textit{1) It a \#Gorgeous day here in \#EmpireStateBuilding...}}  \\
 & \parbox[t]{6cm}{\textit{2) Triceratops Horridus \#threehornedface \#dinosaur...}} \\
 & \parbox[t]{6cm}{\textit{3) New work by \#JoshSmith at @LuhringAugustine...}} \\
 & \parbox[t]{6cm}{\textit{4) Brooklyn. NYC. \#williamsburg \#brooklyn \#nyc ...}} \\
 
 \hline 
 
\multirow{4}{*}{Cluster $1$} & \parbox[t]{6cm}{\textit{1) \#nyccc2015 \#jurassicworld \#NYC \#excited}} \\
 & \parbox[t]{6cm}{\textit{2) Comiccon was everything I expected it to be \#comicconnyc \#comiccon}} \\
 & \parbox[t]{6cm}{\textit{3) \#nycc2015 \#cosplay \#kingpin \#wilsonfisk \#greengoblin}} \\
 & \parbox[t]{6cm}{\textit{4) NYCC Rick Cosplay Meetup 2015!!!! \#ricksanchez \#rickandmorty...}} \\
 \hline 
  
\end{tabular}
\end{center}
\caption{\label{threads:geocluster} Top threads in geo-located clusters at 2015/10/10 17:30 set.}
\end{table}

\begin{figure*}
    \centering
    \includegraphics[width=0.95\textwidth]{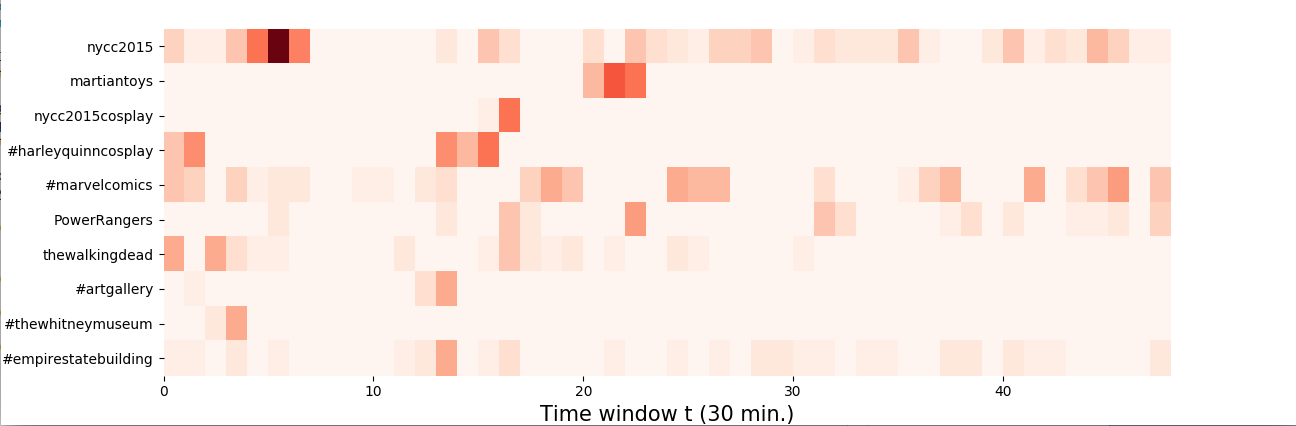}
    \caption{Relevance over time of the top-10 threads in the Comic-con dataset.}
    \label{fig:commicon_rel}
\end{figure*}

Finally, the Threads Ranking module obtained the results depicted in Figure \ref{fig:commicon_rel}, where relevance is calculated each 30 minutes. Automatically, the system detects the anomalous event without any a priori information. The majority of them are related to the Comic-con topics (\textit{Martian Toys}, \textit{Marvel Heroes}, \textit{Power Rangers}, \textit{The Walking Dead premiere}, etc.). However, other threads are completely unrelated to this event such as mentions to exhibitions and important landmarks of the city that people massively visit.

\begin{figure}
    \centering
    \includegraphics[width=0.5\textwidth]{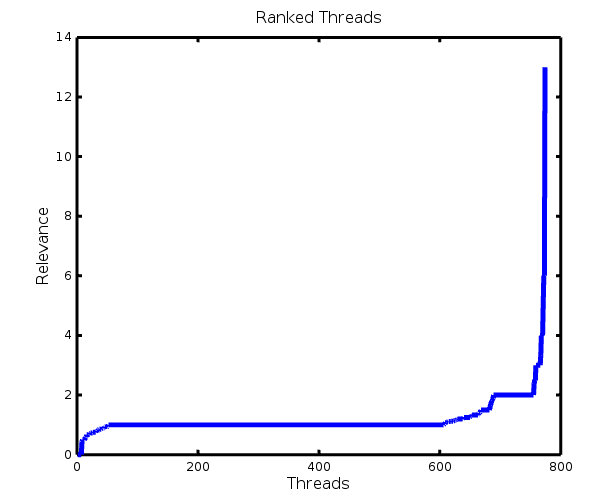}
    \caption{Threads ordered by relevance by the Event Ranking module}
    \label{fig:ordered_threads}
\end{figure}

Using the dataset of Storm Jonas, the discovered threads ranked by their relevance at the end of the day are depicted in Figure \ref{fig:ordered_threads}. As expected, only a few threads are interesting from the geolocation perspective. Pretty curiously, the most relevant thread wasr a Mattise exhibition in the Museum of Modern Arts of New York (MOMA). This can be explained by the absence of big geolocated clusters during this period. In contrast, the biggest content thread in size of this dataset made reference to the Storm Jonas (nearly $1,000$ posts). However, this thread spreads over multiple geo-clusters as it affects the city as a whole. As conclusion, when there is a so large area of influence it is better to check directly the output of the Threads Discovery Module instead of checking the final output.

\subsection{Unexpected Events}
\begin{figure*}
    \centering
    \includegraphics[width=0.95\textwidth]{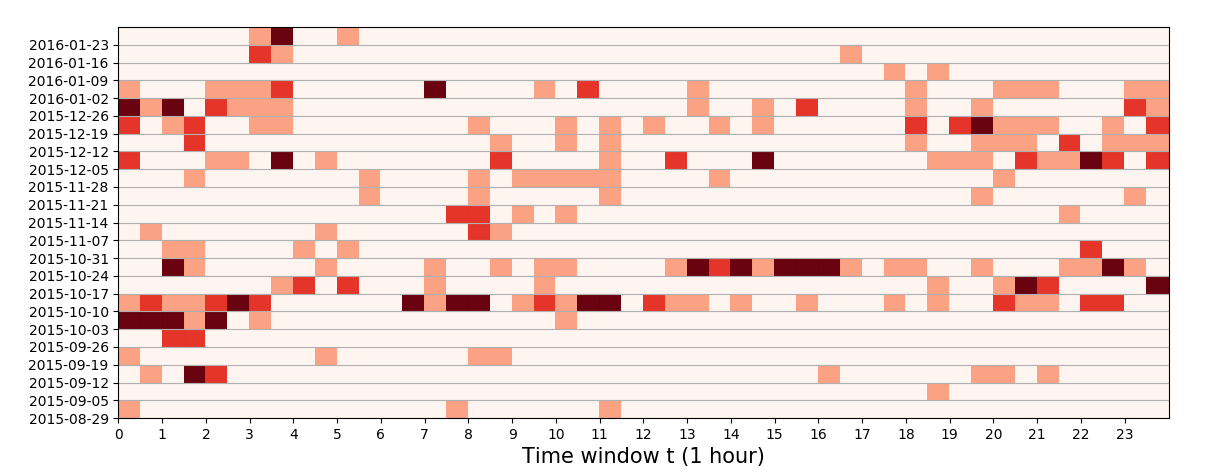}
    \caption{Cluster relevance over time}
    \label{fig:cluster_rel}
\end{figure*}

 As a final test we threw all the data available and we checked the ability of the system in extracting the geolocated clusters with relevant information. Figure \ref{fig:cluster_rel} shows the cluster relevance over time and day. A cluster relevance is obtained as summation of the content relevance of a story at each time step in the cluster. Only the most relevant cluster is depicted at each time interval. Here, a four-scale color intensity portrays the relevance of the thread in comparison with the other threads of that day for easy visualization. As it turns out, highest relevant clusters with uncommon events can be extracted without prior knowledge.

\begin{figure*}
    \centering
    \includegraphics[width=0.95\textwidth]{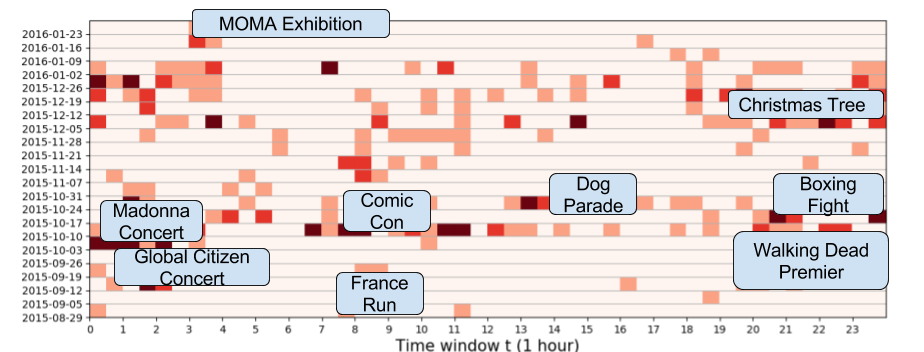}
    \caption{Cluster relevance over time with labelled clusters.}
    \label{fig:cluster_rel_label}
\end{figure*}

Figure \ref{fig:cluster_rel_label} shows some of the important events manually labelled according to the top-3 threads of each cluster. Many of these events are small in absolute terms. For example, both the \textit{Dog Parade} and the \textit{France Run} events are less than the $4\%$ and the $2\%$ of the total amount of posts analyzed that day respectively. Readers must take into account that the posts were gathered coinciding witduring  Christmas time which makes difficult to find rare events not related with this festivity simply by manual observation ($608,492$ different threads are extracted). With our approach, geolocated events easily stand out, drastically lowering the time required by an analyst.

\section{Discussion} 
\label{sec:Discussion}

The experiments in this paper show that combining clustering techniques over GPS-location and text from post shared in LBSNs can effectively discover relevant events which are linked to specific locations in urban areas. The combination of both modules is stronger than each one working on its own. Whilst using only text information the system cannot infer with high precision the locations, although several attempts have been carried out \cite{Schulz:2013,Chandra:2011}, the crowd detection system did not extract the root cause of an event and cannot distinguish posts lying within a certain area but which are not semantically related. 

In spite of the fact that user tags are widely used in social networks such as Instagram, only slightly more than a half of the registers included tags in the caption text ($\approx 56\%$). Thus, naive approaches relying only on tags or certain keywords will filter important data that otherwise will be available for aggregating content. Our approach succeeds in finding relevant content without using specific elements of a social network (e.g. hashtags). Thus, this model could be applied to other scenarios with similar characteristics (e.g. other microblogging services) without extra effort. 

It is also remarkable that a costly infrastructure that must be deployed all over the city, and that normally entails high maintenance costs, is not needed. In fact, the sensing devices (mainly smart phones) are bought, maintained and connected to the Internet by the users. Besides, the whole pipeline of the system is nearly plug and play and it is fed by the users from the posts they publicly share in social media. That entails it could be easily applied to other urban areas only by specifying the configuration parameters: similarity threshold (for the Threads Discovery Module) and $\epsilon$ and $MinPoints$ for the Crowds Dynamics Analyzer. In the first case, the experiments show that, although with slight differences, the aggregated events detected are equivalent for the interval $[0.65, 0.75]$. In the second case, both parameters are easily obtained from the clustering analysis~\cite{dominguez2017sensing}. Furthermore, as the algorithms are fast and low resource consuming, threads resulting from multiple thresholds could be calculated in real time when different granularities are provided. Finally, and since both approaches are unsupervised and can be executed with commodity hardware, this solution opens the door for its application in real scenarios.

\section{Conclusions} 
\label{sec:Conclusions}

Citizens carrying smart devices are turning into sensor on the move. The text and images they actively share in social media, together with the metadata linked to these posts, are definitively an important source of information. In this paper we introduce an Event Detection System that uses geo-located posts to analyze both location and text. The objective is two-folded. On the one hand, detecting unexpected behaviors in the city, i.e. unusual number of people in the same area for the time and day of the week. This task is performed by the Crowds Dynamics Analyzer, which works directly with the GPS-locations. On the other hand, our interest focuses on inferring which the reasons behind these abnormal behaviors are. This task is performed by the Threads Discovery Module. Therefore, we tackled both problems with a combined approach that shares the strengths of a purely mathematical model to detect crowded points in a city, with a Content Aggregation system which is capable of extracting relevant threads of content over vast amounts of data. Our approach is low resource consuming, since all the calculations could be done with commodity hardware which closes the gap for exploitation in real systems.

Our  system does not impose any restriction about the data source if both location and text are provided, although the results are better for short texts. Instagram captions share several characteristics with other social networks (such as Twitter): lack of formality or syntactic structure, misspelled words, new handcrafted tags which glue words or invent new ones. In this scenario, traditional NLP tools for processing input data (e.g. syntactic/dependency parsers, PoS analysis) are more prone to failure. Thus, the shallow clustering technique proposed in this paper is a good alternative to those methods, so that it is more resilient to typos and text informality.

We have assessed this approach over a large dataset of Instagram posts obtained in the New York City area for seven months. The results are promising since it is possible to detect abnormal behaviours in the area as well as being able to infer the reason of that social media activity. Because of these good results and because of the fact no costly infrastructure is needed, this approach might be an additional source of information, helping along with video-surveillance systems in the cities. This is specially the case for Instagram, and the add-ons to include photos in Twitter posts, so that these photos can be taken anywhere (no city infrastructure) and then uploaded to social media. Nevertheless, considering this information as part of a cybersecurity strategy, it should be accompanied by the assessment of credibility, veracity and provenance risks.

Finally, the combination of both modules may enrich the activity patterns obtained by the Crowds Dynamics Analyzer. Nowadays, these patterns only have information about the usual location of citizens all around the urban area at different days of the week and at different times. However, the Threads Discovery Module can add useful information, such as the detection of citizen groups with common interests or the identification of current trends in specific locations of a city. In the current state, the ranking module allows ordering the discovered threads according to their geographical diversity, so prioritizing threads which remain in a neighborhood. As a future work we consider it is also interesting to analyze the geographical and temporal dynamics of the threads, moving from one cluster to another and, and more specifically, the requirements under which a thread in a cluster turns into a multi-hop geolocated thread. According to the classification of rumors as \emph{new} and \emph{long-standing rumors} \cite{zubiaga2017detection}, we are working now in analysing topic, user and time requirements which characterize geo-dependent threads. Our aim is to expand all over the city at the end, by adding geo-location and thread linkage to well-known algorithms for rumor detection \cite{zhang2015and,zubiaga2017towards,Alsaedi:2017:WPR:3068849.2996183}.

\section*{Acknowledgments}
\noindent This work is funded by: the European Regional Development Fund (ERDF) and the Galician Regional Government under agreement for funding  the Atlantic Research Center for Information and Communication Technologies (AtlantTIC), and the Spanish Ministry of Economy and Competitiveness under the National Science Program (TEC2014-54335-C4-3-R and TEC2014-54335-C4-4-R).

\bibliographystyle{ieeetr}
\bibliography{references-gradiant,references-uvigo}



\end{document}